\documentclass[a4paper,12pt]{article}
\usepackage{amsmath,amssymb,bm,fullpage,cite}

\newcommand{\eq}{\begin{equation}}
\newcommand{\qe}{\end{equation}}
\newcommand{\eqa}{\begin{eqnarray}}
\newcommand{\qea}{\end{eqnarray}}
\newcommand{\non}{\nonumber}

\newcommand{\al}{\alpha}
\newcommand{\bt}{\beta}

\newcommand{\ta}{\theta}

\newcommand{\p}{\phi}

\newcommand{\vv}[1]{\ensuremath{\bm{#1}}}

\newcommand{\dd}[2]{\frac{\partial #1}{\partial #2}}
\renewcommand{\d}{\mathrm{d}} 

\newcommand{\ket}[1]{\left| \! \right. #1\left. \! \right\rangle}
\newcommand{\bra}[1]{\left\langle \! \right. #1\left.\! \right| }
\newcommand{\ketbra}[1]{\ket{#1}\bra{#1}}
\newcommand{\braket}[2]{\left\langle \! \right. #1\left.\! \right|\! #2\left. \! \right\rangle}

\newcommand{\comment}[1]{}

\title{\vspace*{-0.1cm}\hfill \raise 20pt \hbox{\small CP3-09-49}\\Noncommuting Coordinates and Magnetic Monopoles}
\author{Jan Govaerts$^{1,2}$\footnote{Fellow of the Stellenbosch Institute for Advanced Study (STIAS), 7600 Stellenbosch, South Africa.}{ }\footnote{jan.govaerts@uclouvain.be}{ } and Se\'an Murray$^{1}$\footnote{sean.murray@uclouvain.be}\\
\\
{\it $^{\rm 1}$Center for Particle Physics and Phenomenology (CP3)}\\
{\it Universit\'e catholique de Louvain}\\
{\it Chemin du Cyclotron 2, B-1348 Louvain-la-Neuve}\\
{\it Belgium }\\
\\
{\it $^{\rm 2}$International Chair in Mathematical Physics and Applications}\\
{\it University of Abomey-Calavi, 072 B. P. 50, Cotonou}\\
{\it Republic of Benin}
}

\date{}
\begin{document}

\maketitle
\begin{abstract}
The appearance of noncommuting spatial coordinates is studied in quantum systems containing a magnetic monopole and under the influence of a radial potential. We derive expressions for the commutators of the coordinates that have been restricted to the lowest energy level. Quantum corrections are found to previous results by Frenkel and Pereira based on quantizing the Dirac brackets of the classical theory. For two different potentials, the modified harmonic oscillator potential and the modified Coulomb potential, we also calculate the commutators for a projection to a fixed energy level.
\end{abstract}

\maketitle

\section*{Introduction}
Over the last decade or so noncommutativity of space coordinates has become a much persued avenue of research \cite{Connes:1994yd}. As well as its use in regularization of quantum field theories, noncommuting coordinates have appeared naturally within string theory \cite{Seiberg:1999vs,Myers:1999ps}. There is also a well-known physical system where such coordinates arise, namely in the motion of a electric particle in an external magnetic field so strong that projection to the lowest Landau (energy) level is justified. The particle is confined to the dimensional space perpendicular to the field and this space becomes noncommutative when the motion is projected to the lowest Landau level.

Suppose the particle has charge $e$ and mass $m$ and is subject to a strong constant magnetic field $\vv{B}$ pointing along the $z-$axis. In the absence of any other external forces the particle is confined to the $(x,y)$ plane. Then it has been shown that after projecting to the lowest Landau level the coordinates satisfy \cite{Dunne:1992ew,Magro}
\eq [x,y]=-i\hbar \frac{c}{eB}~.\qe

In this note, we will generalize this result to the case of a radial magnetic field, namely that of the magnetic monopole\footnote{The case of a particle confined to a sphere centered on a monopole has been considered in \cite{Bander:2004nj}.}. Frenkel and Pereira \cite{Frenkel:2004ff} have made a classical investigation of this problem for strong nonuniform magnetic fields to find non-vanishing Dirac brackets of the coordinates. As we shall see, it is too naive to quantize this system by replacing the Dirac brackets with commutators. We will only carry out the projection after we have quantized the system.

\section*{Magnetic Monopole}
Consider the Hamiltonian for a particle of charge $e$ and mass $m_0$ moving in a magnetic field $\vv{B}=\nabla \times\vv{A}$ in three dimensions $\vv{x}=(x,y,z)=(x_i)$ and under the influence of a radial potential $V(r)$ (in CGS units):
\eq H=\frac{1}{2m_0}(\vv{p}-\frac{e}{c}\vv{A})^2+V(r) ~.\qe
We are interested in the case of a infinitely massive magnetic monopole of magnetic charge $g$ situated at the origin. The corresponding magnetic field is given by
\eq \vv{B}=\nabla \times\vv{A}=g\frac{\vv{x}}{r^3}\qe
and we will take the following choice of vector potential\footnote{The presence of the semi-infinite singularity at $\theta=\pi$, the so-called 'Dirac String' is due to the fact that there does not exist a single analytic vector potential; one must divide the space into (at least) two regions, {\it e.g.} $0\leq\theta\leq \frac{\pi}{2}+\delta$ and $\frac{\pi}{2}-\delta\leq\theta\leq \pi$, and find two smoothly overlapping vector potentials for these regions. Hereafter, we will, where necessary, confine ourselves to the region where the gauge choice given above is analytic. The only true singularity is at the origin and this we remove.}
\begin{align}
A_x&=-\frac{g}{r}\frac{y}{r+z}\qquad A_y=\frac{g}{r}\frac{x}{r+z} \qquad A_z=0~.
\end{align}
\comment{or equivalently
\begin{align}
A_r&=0\qquad A_\theta=0 \qquad A_\phi=\frac{g}{r} \frac{\sin{\theta}}{1+\cos{\theta}}~.
\end{align}}
With this choice, the Schr\"odinger equation becomes
\eq \nabla^2\Psi-\left(\frac{2i\mu}{r^2(1+\cos{\theta})}\dd{}{\p}+\frac{\mu^2}{r^2}\frac{1-\cos{\ta}}{1+\cos{\ta}}+\frac{2m_0}{\hbar^2}V(r)\right)\Psi=-\frac{2m_0}{\hbar^2}E\Psi~,
\qe
where we have introduced the dimensionless quantity $\mu=\frac{eg}{\hbar c}$.
It is well known that the operators
\eq \vv{J}=\frac{1}{\hbar}\vv{x}\times\vv{\pi}-\mu\frac{\vv{x}}{r}~, \qe
where $ \vv{\pi}=\vv{p}-\frac{e}{c}\vv{A}$ are the kinematical momenta, commute with the Hamiltonian and satisfy the $so(3)$ commutation relations\footnote{Some other useful relations are:\eq [{x}_i,\,{\pi}_j]=i\hbar\delta_{ij} \qquad [{\pi}_i,\,{\pi}_j]=i\mu\hbar^2\epsilon_{ijk}\frac{{x}_k}{r^3}\qe and \eq [{J}_i,\,{x}_j]=i\epsilon_{ijk}{x}_k\qquad [{J}_i,\,{\pi}_j]=i\epsilon_{ijk}{\pi}_k~.\qe}
\eq [\hat{J}_i,\,{J}_j]=i\epsilon_{ijk}{J}_k~.\qe
With our gauge choice, we find
\begin{align}
J_x &=i(\sin{\p}\dd{}{\ta}+\cot{\ta}\cos{\p}\dd{}{\p})-\mu\frac{\sin{\ta}\cos{\p}}{1+\cos{\ta}}\\
J_y &=i(-\cos{\p}\dd{}{\ta}+\cot{\ta}\sin{\p}\dd{}{\p})-\mu\frac{\sin{\ta}\sin{\p}}{1+\cos{\ta}}\\
J_z &=-i\dd{}{\p}-\mu
\end{align}

The Schr\"odinger equation can now be rewritten as
\begin{align}
\frac{1}{r}\dd{{}^2}{r^2}(r\Psi)-\frac{1}{r^2}(\vv{J}^2-\mu^2)\Psi-\frac{2m_0}{\hbar^2}V(r)\Psi=-\frac{2m_0}{\hbar^2}E\Psi
\end{align}
This equation is separable and we find
\eq \Psi_{j,m}(r,\ta,\p)=R(r)Y_{\mu,j,m}(\ta,\p)\qe
where $Y_{\mu,j,m}(\ta,\p)$ are the monopole spherical harmonics \cite{Tamm:1931,Fierz:1944,Wu:1976}, which satisfy
\begin{align} \vv{J}^2 Y_{\mu,j,m}&=j(j+1) Y_{\mu,j,m}\\
J_z Y_{\mu,j,m}&=m Y_{\mu,j,m}\\
\mathrm{for~} j&=|\mu|,|\mu|+1,\ldots\qquad m=-j,-j+1,\ldots,j
\end{align}
and $R(r)$ satisfies
\eq \frac{1}{r}\dd{{}^2}{r^2}(rR(r))-\frac{2m_0}{\hbar^2}\left(V-E-\frac{\mu^2\hbar^2}{2m_0 r^2}\right)R(r)=j(j+1)R(r)~.\label{radialeqn}\qe
The monopole spherical harmonics are written explicitly as
\eq Y_{\mu,j,m}(\ta,\p)=M_{\mu,j,m}(1-\cos{\ta})^{(-\mu-m)/2} (1+\cos{\ta})^{(\mu-m)/2} P_{j+m}^{-\mu-m,\mu-m}(\cos{\ta})e^{i(m+\mu)\phi}~,\qe
where $M_{\mu,j,m}=2^m\sqrt{\frac{2j+1}{4\pi}\frac{(j-m)!(j+m)!}{j-\mu)!(j+\mu)!}}$ and $P_n^{\al,\bt}(x)$ are the Jacobi polynomials. When $\mu=0$ the monopole harmonics reduce to the well-known spherical harmonics $Y_{l,m}(\ta,\phi)$.

Dirac showed long ago \cite{Dirac:1931} that in order to have a consistent quantum theory $\mu$ must be quantized. This condition has since been investigated from many different perspectives \cite{Schwinger:1966,Hurst:1968,Peres:1968,Jackiw:1984,Deguchi:2005} and we refer the reader to the literature. For our choice of potential and conventions the quantization condition is
\eq 2\mu\in\mathbb{Z}~.\qe
\paragraph{}
In this note, we shall consider projecting the coordinates to the lowest energy level of the system. We therefore need to specify a potential and we choose potentials that give an energy spectrum which has its minimum at the lowest angular momentum, $j=|\mu|$. The harmonic oscillator and Coulomb potentials fall into this category as do the modified harmonic oscillator and the modified Coulomb potentials, given by
\eq V_{H}(r)=\frac{1}{2}m_0\omega^2 r^2+\frac{\mu^2\hbar^2}{2m_0 r^2}~\quad V_{C}(r)=-\frac{\alpha}{r}+\frac{\mu^2 \hbar^2}{2m_0 r^2}\qe respectively.
The inclusion of the $\frac{1}{r^2}$ term in these latter potentials re-introduces some of the symmetry broken by the monopole: the $SO(4)$ degeneracy group for the Coulomb potential \cite{Zwanziger:1969} and, in the classical case only, the $SU(3)$ symmetry group of the three dimensional harmonic oscillator \cite{Katayama:1993, Yoshida:1989}. For these two cases we will also consider a projection to higher energy levels.

For the modified harmonic potential, the normalised solutions to (\ref{radialeqn}) are just the standard radial functions of the isotropic harmonic oscillator
\eq R^H_{n,j}(r)=\sqrt{\frac{2 n!v^{j+3/2}}{\Gamma(n+j+3/2)}}e^{-\frac{v}{2}r^2}r^j L_n^{(j+1/2)}(vr^2)\quad n\in\mathbb{N}_0\qe
with energy $ E_{N_H}=(N_H+3/2)\omega \hbar $, $N_H=2n+j$ and where $v=\frac{m_0\omega}{\hbar}$. 
Similarly, for the modified Coulomb potential we find the standard functions
\eq R^C_{n,j}(r)=\sqrt{\left( \frac{2}{a N_C}\right)^3\frac{n!}{2N_C(N_C+j)!}}e^{-\frac{r}{a N_C}}\left(\frac{2r}{a N_C}\right)^j L_{n}^{2j+1}\left(\frac{2r}{a N_C}\right)\quad n\in\mathbb{N}_0\qe
for energy $E_{N_C}=-\frac{m_0\al^2}{2\hbar^2 N_C^2}$, $N_C=n+j+1$ and where $a=\frac{\hbar^2}{m_0 \al}$.
We can then write
\eq \Psi^{H,C}_{n,j,m}(r,\ta,\p)=R^{H,C}_{n,j}(r)Y_{\mu,j,m}(\ta,\p)=\braket{r,\ta,\p}{n,j,m}~.\qe

Had we not included the second term in the potentials, the result is essentially the same as above but with $j+1/2$ replaced by $\sqrt{(j+1/2)^2-\mu^2}$. Hence, there is no longer a degeneracy in the energy spectrum since $n$ is still a non-negative integer. As already stated, for the lowest energy projection of the next section, this extra term does not change the structure of the result.

\section*{Projection and Noncommuting coordinates}
Let us consider the projection to the lowest energy level, namely $j=|\mu|$ and $n=0$. The matrix elements of the coordinate operators $\hat{x}_i$ are given by
\begin{align} \bra{m}\hat{x}_i\ket{m'}&=\bra{0,|\mu|,m}\hat{x}_i\ket{0,|\mu|,m'}\non\\&=\int r^2 \sin{\ta}\,\d r\,\d\ta\,\d\p\left(\Psi_{0,|\mu|,m}(r,\ta,\p)\right)^*x_i\Psi_{0,|\mu|,m'}(r,\ta,\p)\end{align}
We can calculate the commutators of these matrices to find after some work
\begin{align} \sum_{m''}\bra{m}\hat{x}_i\ket{m''}\bra{m''}\hat{x}_j\ket{m'}-\sum_{m''}\bra{m}\hat{x}_j\ket{m''}\bra{m''}\hat{x}_i\ket{m'}\non\\=\frac{-i\mathrm{sgn}(\mu)}{|\mu|+1}\,\epsilon_{ijk}\sum_{m''}\bra{m}\hat{r}\ket{m''}\bra{m''}\hat{x}_k\ket{m'}~.\end{align}
Introducing the projector
\eq \mathbf{P}_0=\sum_m\ketbra{0,|\mu|,m} \qe and projected coordinate operators $\hat{X}_i=\mathbf{P}_0\hat{x}_i\mathbf{P}_0$, we can write this as
\eq [\hat{X}_i,\hat{X}_j]=\frac{-i \mathrm{sgn}(\mu)}{|\mu|+1}\,\epsilon_{ijk}\hat{R}\hat{X}_k~,\qe
where $\hat{R}=\mathbf{P}_0\hat{r}\mathbf{P}_0=\int r^3 R_{0,|\mu|}(r)R_{0,|\mu|}(r)\d r\,\mathbf{P}_0$ and since it proportional to $\mathbf{P}_0$, it commutes with the $\hat{X}_i$. 

We see that the projected operators $\hat{X}_i$ are proportional to the usual generators of $SU(2)$
\eq \hat{X}_i=-\frac{\mathrm{sgn}(\mu)}{|\mu|+1}\hat{R}\,\hat{J}_i\qe
and since we are working with the $j=|\mu|$ representation
\eq \sum_{i=1}^3\hat{X}_i\hat{X}_i=\frac{|\mu|}{|\mu|+1}(\hat{R})^2 ~.\qe
Note that $\sum_{i=1}^3\hat{X}_i\hat{X}_i\neq (\hat{R})^2 \neq \hat{R^2}=\mathbf{P}_0\hat{r}^2\mathbf{P}_0$.
	
Notice that in a similar way to the Landau problem, the projected coordinates once again commute as the magnetic charge $g$ (and hence $|\mu|$) goes to infinity. From the above equations we can also see that the projected coordinates are proportional to the matrix coordinates of a fuzzy sphere\cite{Berezin:1974du,Hoppe:1982,Madore:1991bw,Grosse:1993uq,Balachandran:2005ew}, albeit with an unusual normalization. See \cite{Grosse:1994ed,Grosse:1995jt,Presnajder:1999ky} for details of various fields over the fuzzy sphere and of the noncommutative star product. The projected unit vectors $\mathbf{P}_0\frac{\hat{x}_i}{\hat{r}}\mathbf{P}_0$ are more easily seen as fuzzy sphere coordinates, $\hat{R}$ being absent from the analogous equations.

Frenkel and Pereira \cite{Frenkel:2004ff} derived the following commutator of the coordinates $\vv{x}$ by constraining $\vv{\pi}=0$ in the classical Hamiltonian and replacing the Dirac bracket by $\frac{1}{i\hbar}$ times the commutator
\eq [x_i,x_j]=\frac{-i}{\mu}\,\epsilon_{ijk}r x_k~.\qe
Our result includes higher-order terms in $\hbar$
\eq [\hat{X}_i,\hat{X}_j]=\frac{-i}{\mu}\left(1-\frac{1 }{|\mu|}+O(\hbar^2)\right)\,\epsilon_{ijk}\hat{R}\hat{X}_k~.\qe
This is surely the result of a more careful procedure. We have avoided quantizing a constrained singular Hamiltonian by applying the projection after quantization.

\paragraph{Higher Energy Levels}
Let us now consider a projection to some fixed energy level. Since the projection commutes with the angular momentum operators $\hat{J}_i$, the $\hat{X}_i$ are vector operators and so, because of the Wigner-Eckart theorem \cite{Varshalovich}, we expect
\eq \bra{n,j,m}\hat{X}_\varepsilon \ket{n',j',m'}=(-1)^{j-m}\left(\begin{array}{ccc}
j & 1 & j' \\
-m & \varepsilon & m'
\end{array} \right)\bra{n,j}|\hat{X}|\ket{n',j'}~,\label{matrixelement}\qe where $\varepsilon=0,\pm 1$, $\hat{X}_\pm=\hat{X}_1\pm i\hat{X}_2$, $\hat{X}_0=\hat{X}_3$ and the reduced matrix element $\bra{n,j}|\hat{X}|\ket{n',j'}$ is some function which is independent of $m$. We can specify this function by carrying out some integrations. The monopole harmonics satisfy \cite{Wu:1977}
\begin{align} \int \d &\Omega (Y_{\mu,j,m})^* Y_{0,j',m'}Y_{\mu,j'',m''}=\int \d\Omega (-1)^{m+\mu}Y_{-\mu,j,-m}Y_{0,j',m'}Y_{\mu,j'',m''}\non\\
&=(-1)^{m+\mu}(-1)^{j+j'+j''}\sqrt{\frac{(2j+1)(2j'+1)(2j''+1)}{4\pi}}\left(\begin{array}{ccc} j & j' & j'' \\m & m' & m''\end{array} \right)\left(\begin{array}{ccc} j & j' & j'' \\-\mu & 0 & \mu\end{array} \right)\label{monopoleharmonicintegral}\non
\end{align}
where $m+m'+m''=0$.
Then we can use the fact that $Y_{0,1,m}(\ta,\phi)=\frac{1}{2}\sqrt{\frac{3}{\pi}}z$ and $Y_{0,1,\pm1}(\ta,\phi)=\mp \frac{1}{2}\sqrt{\frac{3}{2\pi}}(x\pm i y)=\mp \frac{1}{2}\sqrt{\frac{3}{2\pi}}x_{\pm}$ to provide the angular integrations and give $\bra{n,j}|\hat{X}|\ket{n',j'}$ in term of the Wigner-$3j$ functions and the radial integrations.

For the modified Coulomb potential, we consider coordinates projected to energy level $N_C$,
\eq\hat{X}_i=\mathbf{P}\hat{x}_i \mathbf{P}\quad \mathrm{where}\quad \mathbf{P}=\sum_{j=|\mu|}^{N_c-1}\sum_{m=-j}^j\ketbra{N_C-j-1,j,m}~. \qe
After a long calculation we find that the projected coordinates satisfy the following relation
\eq [\hat{X}_i,\hat{X}_j]=\epsilon_{ijk} \left(a\mu \hat{X}_k+\frac{a^2}{4}(9 N_C^2-\mu^2)\hat{J}_k\right)~. \qe
Hence, $\hat{\vv{X}}$ and $\hat{\vv{J}}$ together form the Lie algebra of $SO(4)$. It is not surprising then that we can write the projected coordinates in terms of the Runge-Lenz vector operator \cite{Zwanziger:1969}
\eq \hat{\vv{A}}=\frac{1}{2m_0}(\hat{\vv{\pi}}\times \hat{\vv{J}}-\hat{\vv{J}}\times\hat{\vv{\pi}})-\al\frac{\hat{\vv{x}}}{r}\qe
as
\eq \hat{\vv{A}}=\frac{4}{3\sqrt{\hbar}}\hat{H}(\hat{\vv{X}}-\frac{a\mu}{2}\hat{\vv{J}})~.\qe
\paragraph{}
Observe that due to (\ref{matrixelement}), $j-j'=0,\pm 1$. Therefore, if we project to a fixed energy, $N_H=2n+j$, for the (un-)modified harmonic potential, only the diagonal terms in the matrix elements are present since either $j=|\mu|,|\mu|+2,\ldots N_H$ or $j=|\mu|+1,|\mu|+3,\ldots N_H$ and so the appropriately projected coordinates satisfy
\eq \hat{X}_i=-\mu \hat{R}\frac{\hat{J}_i}{J^2}\qe and hence
\eq [\hat{X}_i,\hat{X}_j]=-i\mu\,\epsilon_{ijk}\hat{R}\frac{\hat{X}_k}{\hat{J^2}}~\quad\mathrm{and}\quad \sum_{i=1}^3 \hat{X}_i\hat{X}_i=\frac{\mu^2}{J^2}(\hat{R})^2~. \qe
Once again, the projected coordinates commute in the limit $|\mu|\rightarrow\infty$.

\paragraph{Acknowledgments}
This work was supported by the Belgian Federal Office for Scientific, Technical and Cultural Affairs through the Interuniversity Attraction Pole P6/11.

\bibliographystyle{JHEP}

\bibliography{bibfile}

\end{document}